\documentclass[journal]{IEEEtran}
\usepackage{cuted}

\IEEEoverridecommandlockouts

\usepackage{cite}
\usepackage{url}
\usepackage{fancyvrb}
\usepackage[final]{pdfpages}
\usepackage{longtable}
\usepackage{booktabs}
\usepackage[normalem]{ulem}
\usepackage{stfloats}
\usepackage{amsmath}
\usepackage{enumitem}
\usepackage{soul}
\usepackage{amsthm} 
\usepackage{amssymb,bm}
\usepackage[linesnumbered, ruled, boxed]{algorithm2e} 
\usepackage{makecell}
\usepackage{graphicx}
 \newtheorem{theorem}{Theorem} 
   
 \newtheorem{corollary}{Corollary}[theorem]
\newcommand{\nn}{\nonumber}

\ifCLASSINFOpdf
\else
\fi

\begin{document}
\bstctlcite{BSTcontrol}
\setlength{\parindent}{1em}

\title{Source Entropy-Guided Adaptive Transmission for Communication-Driven Multi-View Sensing}

\author{
	Mingjie~Yang,~Guangming~Liang,~\IEEEmembership{Student~Member,~IEEE,} Dongzhu~Liu,~\IEEEmembership{Member,~IEEE,},\\ Lei~Zhang,~\IEEEmembership{Senior~Member,~IEEE}, Xiaonan Liu, and Kaibin Huang, \IEEEmembership{Fellow,~IEEE}
    
    \thanks{(\emph{Corresponding author: Dongzhu Liu}.)}%
	\thanks{M. Yang, G. Liang, and D. Liu are with the School of Computing Science, University of Glasgow, Glasgow G12 8QQ, U.K (e-mail:\{2921021y, 3032221l\}@student.gla.ac.uk; dongzhu.liu@glasgow.ac.uk).}
    \thanks{L. Zhang is with the James Watt School of Engineering, University of Glasgow, Glasgow G12 8QQ, U.K (e-mail: lei.zhang@glasgow.ac.uk).} \thanks{X. Liu is with the School of Natural and Computing Science, University of Aberdeen, AB24 3FX Aberdeen, United Kingdom (email: xiaonan.liu@abdn.ac.uk).}    
    \thanks{K. Huang is with the Department of Electrical and Computer Engineering, The University of Hong Kong, Hong Kong SAR, China (email: huangkb@eee.hku.hk).}      
    \thanks{This paper is an extended version of the conference paper~\cite{YangICC2025} published in IEEE ICC 2025}

}

\markboth{}%
{Shell \MakeLowercase{\textit{et al.}}: Bare Demo of IEEEtran.cls for IEEE Journals}

\maketitle

\begin{abstract}
Communication-driven multi-view sensing relies on routine communication transmissions for sensing acquisition, while the resulting sensing data at distributed devices must be uploaded to an edge server under limited communication resources. This creates a unique coupling between sensing acquisition and edge inference: the communication interval determines the source information, whereas the uplink condition determines how much information can be delivered to the server for sensing inference. To account for this coupling, we propose a source entropy-guided adaptive transmission framework. Specifically, we characterize the entropy of packet-triggered channel state information (CSI) as a function of the communication interval using a multi-output Gaussian process. The resulting analytical bound is compared with the available bit budget, determined by the transmission rate and latency requirement, to select between original-data and task-oriented transmission. For task-oriented transmission, we formulate the communication-constrained inference problem based on the information bottleneck and decompose it into adaptive distributed encoding and multi-view inference (ADE-MI), which avoids alternating optimization between the devices and the edge server.  Experiments on the Widar3.0 multi-view CSI gesture recognition dataset show that the analytical bound closely follows the normalizing-flow numerical estimate, while ADE-MI outperforms task-oriented benchmarks under the same bit budget and the proposed framework further improves recognition accuracy under time-varying channels.

\end{abstract}

%
\IEEEpeerreviewmaketitle

\vspace{-1mm}
\section{Introduction}

Integrated sensing and communication (ISAC) has emerged as a promising paradigm for future wireless systems by enabling sensing and communication to share spectrum, signals, and infrastructure~\cite{9737357,10812728, 9705498, Xing2023,liang2025}. One major research direction in ISAC focuses on the resource tradeoff between sensing and communication, where limited spectrum, transmit power, and spatial degrees of freedom are optimized to balance the performance of the two functions~\cite{9933849,10124135,9916163,10556683}. In addition to this widely studied resource-sharing perspective, another research direction, which is the focus of this work, explores the synergy between communication and sensing, where communication signals are reused for sensing without dedicated transmission design ~\cite{10.1145/3570361.3613274,9140020}. WiFi sensing is a representative example of this paradigm, as channel state information (CSI) extracted from routine communication packets can capture environmental dynamics and support applications such as human activity and gesture recognition~\cite{10.1145/3310194,10.1145/2789168.2790093}.

In this passive sensing paradigm, the sensing capability depends on the communication opportunities arising from routine transmissions. Specifically, lower packet transmission rates or longer channel-sampling intervals reduce the temporal density of CSI observations and can degrade 
sensing accuracy~\cite{10192291}. Under practical irregular CSI acquisition, sensing performance depends not only on the number of available samples, but also on when they are collected and how fresh they are at the time of inference~\cite{zakeri2026}.  However, prior studies characterize these effects in terms of downstream sensing performance, without quantifying how changes in the communication interval affect the information content of the resulting sensing data. Such information content becomes particularly important for edge inference, where sensing data collected at distributed devices must be transmitted to an edge server and jointly processed to support sensing tasks under limited communication resources.

Task-oriented communication addresses the communication bottleneck of edge inference by transmitting compact representations that preserve only task-relevant information~\cite{9955582,10183789,10186369,Zhu2020}. This introduces a tradeoff between inference performance and representation compression, which can be formally characterized by the information bottleneck (IB)~\cite{tishby2000information,aguerri2019distributed}. However, IB-based multi-device cooperative inference schemes typically treat the sensing data as already acquired and optimize task-relevant representations under a given communication constraint~\cite{9837474,9606667}. Such designs do not capture the communication-dependent nature of passive sensing, where the communication interval changes the amount and information content of the acquired sensing data, while the uplink conditions determine the channel capacity available for edge inference, leading to a potential mismatch between source information demand and available transmission rate.

To address this mismatch, we propose a source entropy-guided adaptive transmission framework that jointly accounts for communication-dependent sensing acquisition and time-varying uplink conditions for edge inference.  Unlike IB-based task-oriented communication~\cite{9606667,10159007}, we explicitly characterize the source entropy as a function of the communication interval and use it together with the available channel capacity to determine whether the sensing data should be transmitted in their original form or through task-oriented encoding.

\subsection{Related Work} \label{sec:related_work} 
    \subsubsection{Communication-Driven Passive Sensing} 
    Prior studies have examined how the temporal density of CSI acquisition affects sensing  by varying CSI acquisition intervals~\cite{10192291} or packet generation rates~\cite{9918170}. However, sensing accuracy depends not only on the number of acquired CSI frames, but also on when they are observed and how fresh they are at the time of inference~\cite{zakeri2026}. Similar observations have been shown in the frequency domain: increasing bandwidth or the number of subchannels does not necessarily yield higher sensing gains, whereas finer spectral resolution can provide more informative channel observations~\cite{10192291,10099368}. On the other hand, spatial diversity through additional antennas or distributed receivers can further improve inference performance~\cite{9842828}. These results suggest that simply increasing the number or dimensionality of sensing observations does not necessarily provide proportional  gain. This motivates a direct characterization of how communication-driven acquisition changes the source information contained in the resulting CSI.

    

	\subsubsection{Task-Oriented Communication}

Existing task-oriented communication primarily focuses on how to efficiently transmit sensing data after they have been acquired. Information-bottleneck (IB) methods provide a principled way to learn compact task-relevant representations~\cite{tishby2000information}, and distributed IB further extends this idea to multi-device cooperative inference~\cite{aguerri2019distributed,9837474}. More recent studies introduce channel and latency awareness by dynamically adapting the amount of transmitted task information, including feature-dimension control for robustness under poor channels~\cite{9606667}, semantic-symbol adaptation to input and channel conditions~\cite{10621397}, joint compression and resource allocation under latency constraints~\cite{10375768}, and progressive feature transmission until the additional inference gain becomes limited~\cite{9955582}.
At the sensing-system level, Ultra-LoLa further considers the tradeoff between packet reliability and the number of sensing observations that can be delivered within a latency deadline~\cite{wang2025ultra}.  However, these approaches overlook how communication affects source acquisition and thus the information available for edge inference. 

	\subsubsection{Entropy Characterization of Sensing Data}
    Quantifying source information in high-dimensional sensing data is challenging because entropy estimation requires the unknown joint distribution of the observations. Non-parametric estimators such as k-NN methods~\cite{kraskov2004estimating} become increasingly data-demanding as dimensionality grows~\cite{10.5555/3295222.3295348}. Normalizing flows alleviate this limitation by learning flexible parametric density models for complex high-dimensional distributions~\cite{papamakarios2021}. However, such numerical estimators fail to show how the communication interval changes the source information of the resulting CSI, motivating us to develop an analytical approach and use a normalizing flow for numerical validation.

\subsection{Contributions}

This work considers multi-view passive sensing, where distributed devices acquire sensing observations from routine communication packets and subsequently upload them to an edge server for joint sensing inference. We investigate the coupling between these two communication stages: the communication interval determines the sensing information acquired at the devices, while the uplink condition constrains how much of this information can be delivered to the edge server.  We characterize this communication-dependent source information and use it to develop an adaptive transmission framework for multi-view sensing inference. The main contributions and findings are summarized as follows. 

\noindent $\bullet$ \textbf{Entropy Analysis of Communication-Dependent Sensing Data:} We characterize the entropy of packet-triggered CSI sensing data as a function of the communication interval \(\Delta\)  using a multi-output Gaussian process (MOGP) and derive a tractable analytical upper bound which shows an \(O(1/\Delta)\) scaling law in the long-interval regime. The analytical result is further validated against a numerical entropy estimate via a normalizing-flow.  
    
\noindent $\bullet$ \textbf{Source entropy-guided adaptive multi-view transmission and inference:}  We formulate a communication-constrained multi-view inference problem, where the available transmission rate and latency requirement determine the bit budget. The source entropy is compared with this budget to select between original-data transmission and task-oriented transmission. For task-oriented transmission, we decompose the problem into adaptive distributed encoding (ADE) and multi-view inference (MI), which avoids alternating optimization between the devices and the edge server. The  MI  also applies when the original sensing data are transmitted. 

\noindent$\bullet$ \textbf{Experiments:} Experiments on Widar3.0 for multi-view CSI gesture recognition show that the analytical entropy bound closely follows the normalizing-flow numerical estimate.
For task-oriented transmission, ADE-MI outperforms the VDDIB benchmark~\cite{9837474} under the same bit budget, while the  adaptive transmission framework further improves recognition accuracy under time-varying channels.

\begin{figure*}[t!]
    \centering
    \includegraphics[scale=0.55]{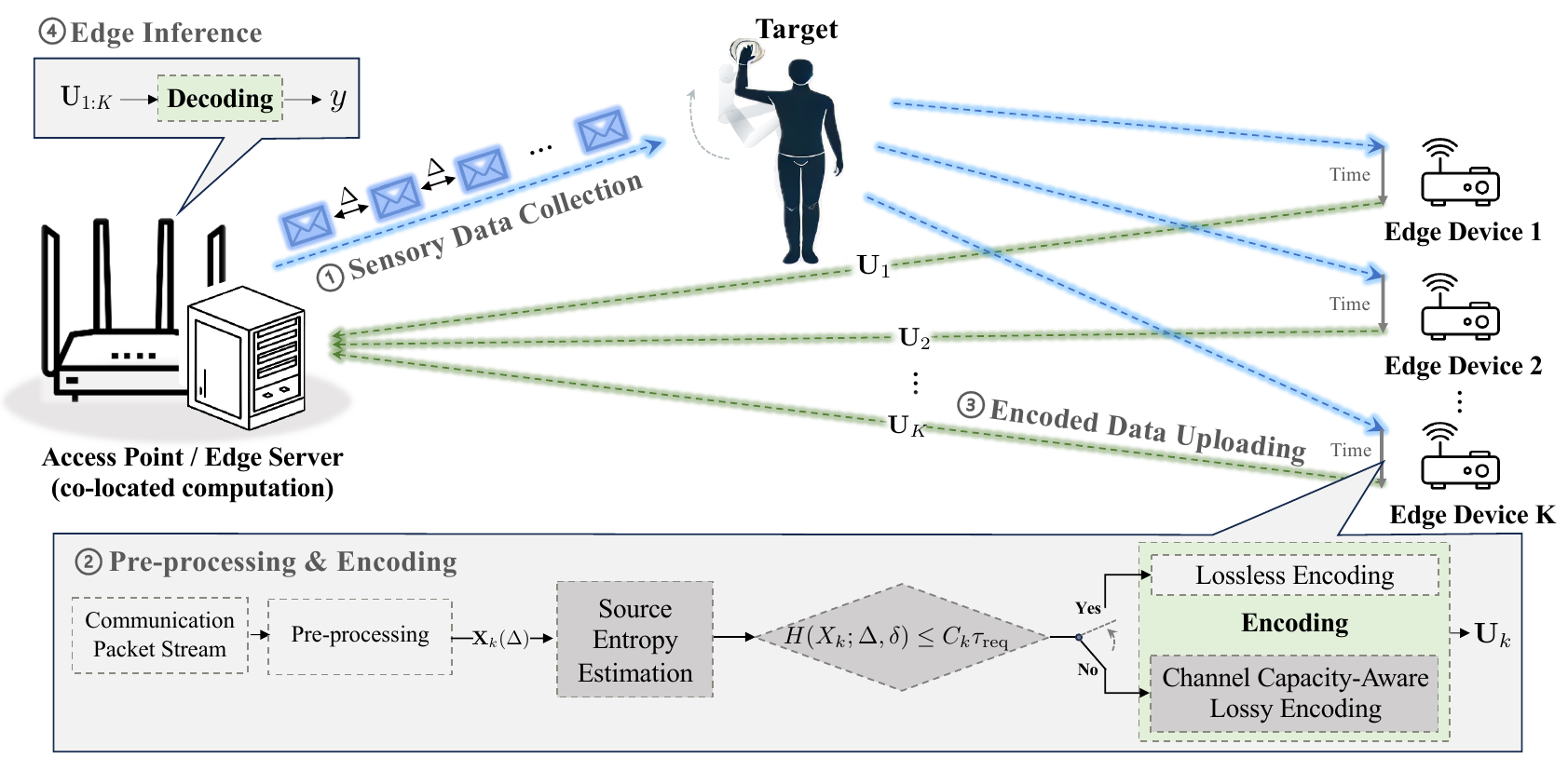}
    \caption{Overview of the proposed framework for adaptive distributed sensing and inference.}
    \label{fig:model1}
\end{figure*}

\section{System Model}\label{Sec1}
	
	\subsection{Sensing Model} \label{Sec2subsec1}
We consider a multi-view WiFi sensing system as illustrated in Fig.~\ref{fig:model1}, consisting of one access point (AP) and $K$ edge devices. 
The system monitors human activities over finite observation windows. As a representative application, we consider gesture recognition, where each sensing event corresponds to one gesture instance performed within an observation window of duration $\tau$, with class-label 
$y$.

In this passive WiFi sensing system, sensing observations are extracted from routine communication packets. Specifically, each received packet yields a CSI snapshot via channel estimation, which reflects the instantaneous propagation environment.
The sequence of CSI snapshots collected over the observation window $\tau$ forms the raw sensing observation for one event. Consequently, the temporal resolution of sensing is determined by the packet arrival process.
For analytical tractability, we characterize the packet stream by an effective communication interval $\Delta$, defined as the average inter-packet interval over the observation window $\tau$. For irregular real-world traffic, $\Delta$ can be interpreted as a local average over short time periods during which packet arrivals are approximately time-invariant.

After standard preprocessing, such as amplitude extraction or dimensionality reduction, the sensing data collected at device $k$ is represented by an effective CSI matrix
\[
\mathbf{X}_k \in \mathbb{R}^{T(\Delta)\times D},
\]
whose entries are CSI features after finite-resolution hardware quantization with resolution $\delta$. This hardware-induced quantization is distinct from the transmission-oriented quantization design considered later.
Here $T(\Delta) = \lfloor \tau/\Delta \rfloor$ denotes the number of received packets within the observation window, and $D$ is the feature dimension of each preprocessed CSI snapshot.
Therefore, the continuous physical process is represented digitally through temporal sampling with interval $\Delta$ and quantization with resolution $\delta$. Together with the observation duration $\tau$, these parameters determine the amount of information contained in the sensing data.

\subsection{Communication Model for Edge Inference}

We consider an access-side edge deployment where the edge server is co-located with the AP. As a result, the delay between the AP and the edge server is negligible, and the overall communication latency is dominated by the wireless uplink from edge devices.
 
Each device uploads its local sensing information to the edge server through an OFDMA uplink for edge inference. Let $B$ denote the total bandwidth and $B_k = B/K$ as bandwidth allocated to device $k$. $\gamma_k$ denotes its received signal-to-noise ratio (SNR). The achievable transmission rate of device $k$ is
\[
C_k = \zeta B_k \log_2(1+\gamma_k),
\]
where $\zeta \in (0,1]$ accounts for practical losses due to modulation, coding, and protocol overhead. Given a latency requirement $\tau_{\mathrm{req}}$, the maximum number of bits that device $k$ can transmit is $C_k \tau_{\mathrm{req}}$.

The sensing data matrix $\mathbf{X}_k$ collected within one observation window is regarded as a realization of a discrete random variable $X_k$. Let $H(X_k;\Delta, \delta)$ denote the Shannon entropy of $X_k$, which depends on the effective communication interval $\Delta$ and the quantization resolution $\delta$. 
This quantity characterizes the minimum average number of bits required for lossless source coding. 

Let $\mathbf{U}_k$ denote the uploaded data generated from $\mathbf{X}_k$ at device $k$. 
Under the latency constraint, original-data transmission of the raw sensing data is feasible if
\begin{align}\label{eq:lossless_trans}
H(X_k;\Delta, \delta) \le C_k\tau_{\mathrm{req}}.
\end{align}
When \eqref{eq:lossless_trans} holds, the edge server can reconstruct $\mathbf{X}_k$ from $\mathbf{U}_k$  without distortion.

Otherwise, device $k$ encodes $\mathbf{X}_k$ into a task-oriented low-dimensional representation $\mathbf{Z}_k \in \mathbb{R}^{d_k}$, which serves as the uploaded data, i.e., $\mathbf{U}_k=\mathbf{Z}_k$.  Let $Z_k$ denote the random variable corresponding to $\mathbf{Z}_k$. 
From an information-theoretic perspective, the communication cost of the encoded representation is characterized by the mutual information between $X_k$ and $Z_k$, which should satisfy
\begin{align}\label{eq:lossy_trans}
I(X_k;Z_k) \le C_k\tau_{\mathrm{req}}.
\end{align}


\subsection{Communication-Constrained Inference Model}

We adopt a probabilistic learning framework for multi-view cooperative edge inference, where $Y$ denotes the random variable corresponding to the target label $y$.
When the condition in \eqref{eq:lossless_trans} is satisfied for all devices, the edge server can reconstruct the raw sensing data 
$\{\mathbf{X}_k\}_{k=1}^{K}$ and then directly perform inference according to the conditional distribution 
$p_{\phi}(y|\mathbf{X}_1,\ldots,\mathbf{X}_K)$, where $\phi$ denotes the parameters of the raw-data inference model.

Otherwise, the encoding process at device $k$ is modeled by the conditional distribution
$p_{\theta_k}(\mathbf{Z}_k|\mathbf{X}_k)$, where $\theta_k$ denotes the encoder parameters. The corresponding random variables satisfy the Markov chain
\begin{align}
    Y \leftrightarrow X_k \leftrightarrow Z_k, \quad \forall k \in \{1, \ldots, K\},
\end{align}
which implies that $Z_k$ is conditionally independent of $Y$ given $X_k$ for each device $k$.  Based on the received representations $\{{\bf Z}_1, \cdots, {\bf Z}_K\}$, the edge server performs inference according to
$p_{\psi}(y|\mathbf{Z}_1,\ldots,\mathbf{Z}_K)$, parameterized by $\psi$.


The design objective is to maximize inference performance under communication constraints. This introduces a fundamental tradeoff between communication efficiency and the preservation of task-relevant information, which will be formalized and further characterized in the following sections. The above formulation adopts a common transmission mode across devices, which provides a concise framework for analyzing when original data or task-oriented features should be transmitted and how the edge server performs inference in each case. Nevertheless, the inference model can be extended to the case where each device independently selects its transmission mode, as discussed in the future work.

\section{Entropy Analysis of Sensing Data}  \label{Sec2}

As introduced in the system model, the entropy of the quantized sensing data $H(X_k;\Delta,\delta)$ serves as the decision criterion in the proposed adaptive transmission framework. However, direct evaluation of this quantity is challenging because ${\bf X}_k$ is a high-dimensional quantized sensing matrix with an extremely large discrete state space. As a result, the corresponding discrete distribution cannot be reliably estimated from empirical frequencies. To address this challenge, we exploit the physical data-generation mechanism described in Sec.~\ref{Sec2subsec1} 
and relate the discrete entropy of the quantized sensing source to the differential entropy of its unquantized counterpart. 
When the quantization step $\delta$ is sufficiently small relative to the variations represented in the sampled CSI data, the classical high-resolution relation between discrete entropy and differential entropy yields~\cite{Cover2006,GrayNeuhoff1998}
\begin{equation}
    H(X_k;\Delta,\delta) \approx h(\tilde{X}_k;\Delta) - T(\Delta)D \ln(\delta),
    \label{eq:h_to_H}
\end{equation}
where $\widetilde{X}_k$ denotes the unquantized sampled CSI, $T(\Delta)$ is the number of temporal samples, and $D$ is the dimension of each sample. For each fixed communication interval $\Delta$, this relation allows the entropy of the quantized sensing data to be approximated through the differential entropy of its unquantized counterpart.


 Despite this transformation, characterizing 
$h(\tilde{X}_k;\Delta)$ remains nontrivial because the underlying 
continuous distribution of the CSI observations is unknown. Nevertheless, the CSI observations exhibit structural temporal and cross-feature dependencies that can be used for analytical distribution modeling.
We therefore first develop an analytical approach 
that exploits this structure and derives a tractable upper-bound 
approximation using a multi-output Gaussian process (MOGP) model. 
Besides quantifying the source information volume, the resulting 
expression also reveals an interpretable scaling law in the 
sparse-sampling regime. We then employ normalizing flows (NF) to 
obtain a data-driven numerical estimate of the entropy and evaluate 
the tightness of the analytical bound.

 \subsection{MOGP-Based Analytical Upper Bound on Sensing Entropy}
\label{subsec:mogp_bound}

To obtain a tractable analytical bound on the sensing entropy 
$H(X_k;\Delta,\delta)$, we first characterize the differential 
entropy of the unquantized sampled CSI process. Let 
$X_k(t)\in\mathbb{R}^{D}$ denote the continuous-time CSI process of device $k$. Given the communication 
interval $\Delta$, the unquantized sensing matrix 
$\tilde{X}_k\in\mathbb{R}^{T(\Delta)\times D}$ is obtained by 
sampling $X_k(t)$ at $t=i\Delta$, $i=1,\ldots,T(\Delta)$, with the 
$i$-th row given by $X_k(i\Delta)$. Thus, characterizing 
$h(\tilde{X}_k;\Delta)$ amounts to modeling the joint distribution 
of these sampled CSI vectors.

Since this joint distribution is unknown and difficult to estimate 
directly, we model $X_k(t)$ as a multi-output Gaussian process 
(MOGP). This model induces a Gaussian surrogate for the vectorized 
matrix $\mathrm{vec}(\tilde{X}_k)$, with covariance determined by 
the sampled covariance function of $X_k(t)$. 
By the maximum entropy 
property of the Gaussian distribution, among all distributions 
sharing the same covariance matrix, this surrogate provides an 
upper bound on the true differential entropy~\cite{Cover2006}:
\begin{equation}
    h(\tilde{X}_k;\Delta)
    \leq
    h_{\mathrm{GP}}(\tilde{X}_k;\Delta).
    \label{eq:gp_entropy_bound}
\end{equation}
This yields a conservative transmission-control criterion: satisfying 
the channel-capacity constraint in~\eqref{eq:lossless_trans} with the 
entropy upper bound also guarantees feasibility for the true entropy.

We now specify the MOGP model for $X_k(t)$ to compute 
$h_{\mathrm{GP}}(\tilde{X}_k;\Delta)$. The process 
$X_k(t)$ is modeled as a zero-mean MOGP 
\begin{equation}
    X_k(t) \sim \mathcal{GP}\!\left(\mathbf{0},
    \mathbf{C}(t,t')\right),
\end{equation}
where $\mathbf{C}(t,t')\in\mathbb{R}^{D\times D}$ is covariance function. 
Motivated by the fact that CSI variations arise from both temporal 
dynamics and structured correlations across CSI features, we adopt 
the separable covariance model~\cite{alvarez2012kernels}
\begin{equation}
    \mathbf{C}(t,t')
    =
    k_{\mathrm{time}}(t,t')\mathbf{K}_d,
    \label{eq:separable_cov}
\end{equation}
where $k_{\mathrm{time}}(t,t')$ captures the correlation across time, 
and $\mathbf{K}_d\in\mathbb{R}^{D\times D}$ captures the static 
feature-domain correlation induced by multipath propagation, antenna 
configuration, and subcarrier correlation.
For the temporal kernel, we use the rational quadratic (RQ) kernel
\begin{equation}
    k_{\mathrm{time}}(t,t')
    = \sigma^2 \left( 1+ \frac{(t-t')^2}{2\alpha\ell^2} \right)^{-\alpha}, \label{eq:rq_kernel}
\end{equation}
where $\sigma^2$, $\ell$, and $\alpha$ denote the signal variance, 
length scale, and scale-mixture parameter, respectively. These 
parameters, together with $\mathbf{K}_d$, are estimated from the 
collected data and kept fixed when evaluating the entropy bound.

Accordingly, sampling the MOGP at $\{i\Delta\}_{i=1}^{T(\Delta)}$ 
yields a multivariate Gaussian distribution for the vectorized 
unquantized sensing matrix:
\begin{equation}
    \mathrm{vec}(\tilde{X}_k)
    \sim
    \mathcal{N}\left(
    \mathbf{0},
    \mathbf{K}_t(\Delta)\otimes\mathbf{K}_d
    \right),
\end{equation}
where $\mathbf{K}_t(\Delta)\in\mathbb{R}^{T(\Delta)\times T(\Delta)}$ 
is obtained by evaluating $k_{\mathrm{time}}(t,t')$ at the sampled 
time instants, i.e., its $(i,j)$-th entry is 
$k_{\mathrm{time}}(i\Delta,j\Delta)$.

The MOGP-induced differential entropy is then given by the entropy 
of a multivariate Gaussian:
\begin{equation}
\begin{aligned}
    h_{\mathrm{GP}}(\tilde{X}_k;\Delta)
    &=
    \frac{T(\Delta)D}{2}\ln(2\pi e)
    +
    \frac{D}{2}\ln\det(\mathbf{K}_t(\Delta)) \\
    &\quad+
    \frac{T(\Delta)}{2}\ln\det(\mathbf{K}_d).
\end{aligned}
\label{eq:gp_entropy}
\end{equation}
Substituting \eqref{eq:gp_entropy} into the quantization relation 
in~\eqref{eq:h_to_H} yields the following theorem.

\begin{theorem}[Analytical Upper Bound on Sensing Entropy]
    \label{thm:main_entropy_bound}
    Under the MOGP covariance model in~\eqref{eq:separable_cov} 
    and within the high-resolution quantization regime 
    characterized by~\eqref{eq:h_to_H}, the entropy of the 
    quantized sensing data $H(X_k;\Delta,\delta)$ admits the 
    following analytical upper-bound approximation:
    \begin{equation}
    \begin{aligned}
        H(X_k;\Delta,\delta)
        &\lesssim
        H_{\mathrm{ub}}(X_k;\Delta,\delta) \\
        &=
        \frac{T(\Delta)D}{2}\ln(2\pi e)
        +
        \frac{T(\Delta)}{2}\ln\det(\mathbf{K}_d)  \\
        &\quad+
        \frac{D}{2}\ln\det(\mathbf{K}_t(\Delta))
        - T(\Delta)D\ln(\delta).
        \label{eq:main_bound}
    \end{aligned}
    \end{equation}
\end{theorem}
{\color{blue} }
The result in~\eqref{eq:main_bound} separates the effects of the 
sampling volume, feature-domain covariance, temporal covariance, 
and quantization resolution. The first term is the 
dimension-dependent constant in the Gaussian entropy formula. 
The second and third terms characterize the effects of 
feature-domain and temporal dependencies through 
$\mathbf{K}_d$ and $\mathbf{K}_t(\Delta)$, respectively. 
If these matrices are normalized as correlation matrices, their 
log-determinants quantify the entropy reduction induced by 
redundancy. In the general covariance case, the log-determinants 
jointly reflect both statistical dependence and variance scaling. 
The last term captures the increase in source entropy caused by finer quantization.

 The entropy bound in~\eqref{eq:main_bound} relies on estimating the feature covariance and temporal kernel parameters, including $\sigma^2$, $\ell$, $\alpha$, and $\mathbf{K}_d$, from sensing data. These parameters capture the statistics of the sensing environment and the monitored activities. Once fitted under a given operating condition, they can be used to evaluate the entropy upper-bound approximation in closed form for different communication intervals $\Delta$, without repeated parameter estimation. If the environment or activity statistics change, the parameters can be re-estimated from calibration or recent sensing data and substituted into the same entropy-bound expression.


The communication interval $\Delta$ affects the bound through two 
coupled mechanisms. Decreasing $\Delta$ increases the number of 
collected samples, $T(\Delta)=\left\lfloor\frac{\tau}{\Delta}\right\rfloor$,
but also strengthens the temporal dependence between adjacent 
samples through $\mathbf{K}_t(\Delta)$. Hence, the bound captures 
the trade-off between the increased sampling volume and the 
temporal redundancy introduced by more frequent transmissions.

This coupling also provides insight into the limiting behavior of the 
bound. In the long-interval regime, where $\Delta\gg\ell$ for the RQ 
kernel in~\eqref{eq:rq_kernel}, the temporal covariance between distinct 
packet-induced CSI snapshots becomes negligible. Consequently, 
$\mathbf{K}_t(\Delta)$ becomes approximately diagonal:
\begin{equation}\label{eq:diag_Kt}
    \mathbf{K}_t(\Delta)
    \approx
    \sigma^2\mathbf{I}_{T(\Delta)}.
\end{equation}
Applying~\eqref{eq:diag_Kt} to~\eqref{eq:main_bound} yields the following 
corollary.


\begin{corollary}[Sparse-Sampling Scaling of the Sensing Entropy Bound]
    \label{cor:sparse_sampling_scaling}
    When the communication interval is much larger than the 
    temporal length scale of the sensing process, i.e., 
    $\Delta\gg\ell$, the entropy upper-bound approximation satisfies
    \begin{equation}
    \begin{aligned}
        H_{\mathrm{ub}}(X_k;\Delta,\delta)
        &\approx
        T(\Delta)
        \Bigg[
        \frac{D}{2}\ln(2\pi e)
        +
        \frac{1}{2}\ln\det(\mathbf{K}_d) \\
        &\qquad\qquad
        +
        \frac{D}{2}\ln\sigma^2
        -
        D\ln\delta
        \Bigg].
        \label{eq:sparse_sampling_entropy}
    \end{aligned}
    \end{equation}
    Since 
    $T(\Delta)=\lfloor\tau/\Delta\rfloor$, the bound is 
    approximately proportional to $1/\Delta$ when the integer 
    rounding effect is negligible:
    \begin{equation}
        H_{\mathrm{ub}}(X_k;\Delta,\delta)
        \big|_{\Delta\gg\ell}
        =
        \mathcal{O}\!\left(\frac{1}{\Delta}\right).
        \label{eq:sparse_sampling_scaling}
    \end{equation}
\end{corollary}


Corollary~\ref{cor:sparse_sampling_scaling} shows that, in the 
sparse-sampling regime, the entropy upper bound is dominated by the 
number of packet-induced CSI snapshots.
Since $\Delta\gg\ell$ makes 
adjacent snapshots nearly uncorrelated, each snapshot contributes an 
approximately constant amount of information, and the entropy 
scales as $\mathcal{O}({1}/{\Delta})$. 
This inverse scaling clarifies the behavior of the original-data-transmission 
condition in~\eqref{eq:lossless_trans}: for a fixed uplink budget 
$C_k\tau_{\mathrm{req}}$, increasing $\Delta$ reduces the sensing entropy, so original data upload becomes feasible once the 
entropy bound falls below the available transmission budget.

 \subsection{Numerical Validation via Normalizing Flows}
\label{subsec:nf_validation}

The MOGP-based result in Sec.~\ref{subsec:mogp_bound} yields a closed-form upper bound on the sensing entropy by using a Gaussian surrogate for the sampled CSI distribution. Since the resulting bound relies on the Gaussian assumption, we validate its tightness using normalizing flows (NFs) as an offline numerical entropy estimator. Unlike the MOGP model, the NF does not impose a Gaussian distribution on the CSI samples. Instead, it represents the unknown CSI distribution as the pushforward of a tractable base distribution, typically a standard Gaussian, through an invertible transformation.

Specifically, in our settings, the NF learns a distribution $p_{\theta}$ to approximate the unquantized CSI distribution ${p}_{\tilde{X}_k}$, where $p_{\theta}$ is induced 
by pushing forward a standard Gaussian base distribution through an invertible mapping $g_{\theta}(\cdot)$.  The objective 
of NF training can be formulated as

\begin{equation}
    \theta^{\star}
    =
    \arg\min_{\theta}
    D_{\mathrm{KL}}\!\left(
    {p}_{\tilde{X}_k}\|p_{\theta}
    \right).
    \label{eq:nf_kl}
\end{equation} 
In practice, this objective is implemented by maximizing the likelihood of the collected CSI samples under $p_{\theta}$ \cite{Rezende2015,papamakarios2021}.

After training, the entropy of the NF-induced distribution can be computed via the change-of-variables formula, 
\begin{equation}
\label{eq:nf_entropy_identity}
    h_{\theta}(\tilde{X}_k;\Delta)
    =
    \frac{T(\Delta)D}{2}\ln(2\pi e)
    +
    \mathbb{E}_{\mathbf{u}\sim\mathcal{N}(\mathbf{0},\mathbf{I})}
    \left[
    \ln\left|\det J_{g_{\theta}}(\mathbf{u})\right|
    \right],
\end{equation}
where $\mathbf{u}$ denotes the latent variable drawn from the standard Gaussian base distribution, and $J_{g_{\theta}}(\mathbf{u})$ is the Jacobian matrix of $g_{\theta}$ at $\mathbf{u}$. The first term is the entropy of the standard Gaussian base distribution, while the second term measures the expected log-volume change induced by the flow when transforming the base distribution into the CSI distribution. Therefore, $h_{\theta}(\tilde{X}_k;\Delta)$ serves as an offline numerical estimate of the differential entropy $h(\tilde{X}_k;\Delta)$.

We note that the NF estimate is not used as the online transmission-control metric, but rather as an offline numerical reference for assessing the analytical MOGP bound. This is because each communication interval $\Delta$ leads to a different input dimension and therefore requires training a separate high-dimensional flow model, which is computationally expensive. Furthermore, evaluating the NF entropy requires neural-network inference and Jacobian log-determinant computation, which are unsuitable for real-time decision making under the latency requirement $\tau_{\mathrm{req}}$. By contrast, once the MOGP covariance parameters are estimated offline, the upper bound of $H(X_k;\Delta,\delta)$ in \eqref{eq:main_bound} can be evaluated in closed form. The comparison between the NF-based numerical estimate and the analytical MOGP bound is presented in Sec.~\ref{Sec_exp}, Fig.~\ref{fig:source_entropy}.

\section{Source Entropy-Driven Adaptive Transmission for Multi-View Inference}
\label{Sec4}

Our objective is to maximize inference accuracy at the edge server under the communication constraint. To achieve this, the information received by the server $U=\{U_1,\cdots,U_K\}$ should jointly preserve as much information about $Y$ as possible, which can be measured by the mutual information $\mathrm{I}(Y;U)$. Since $\mathrm{I}(Y;U)=\mathrm{H}(Y)-\mathrm{H}(Y|U)$ and $\mathrm{H}(Y)$ is constant, maximizing $\mathrm{I}(Y;U)$ is equivalent to minimizing $\mathrm{H}(Y|U)$. Therefore, the communication-constrained multi-view inference problem is formulated as

\begin{equation*}
\textbf{(P0)} \quad
\begin{aligned} \label{eq:opt0}
    &\min \quad \mathrm{H}(Y|U_1,U_2,\cdots,U_K) \\
    &\mathrm{s.t.} \quad \mathrm{I}(X_k;U_k)\leq C_k\tau_{\mathrm{req}},\quad \forall k\in\{1,2,\ldots,K\}.
\end{aligned}
\end{equation*}

The source-entropy bound analyzed in Sec.~\ref{Sec2} provides a tractable 
criterion for instantiating the optimization problem~\textbf{P0}. 
If the original-data-transmission condition
$H_{\mathrm{ub}}(X_k;\Delta,\delta) \leq C_k\tau_{\mathrm{req}}$
is not satisfied for at least one device, the received information 
$U_k$ is set to the low-dimensional task-oriented representation $Z_k$. 
In this case, \textbf{P0} becomes
\begin{equation}
\textbf{(P1)} \quad
\begin{aligned}
    &\min \quad \mathrm{H}(Y|Z_1,Z_2,\ldots,Z_K) \\
    &\mathrm{s.t.} \quad 
    \mathrm{I}(X_k;Z_k)\leq C_k\tau_{\mathrm{req}},
    \quad \forall k\in\{1,2,\ldots,K\}.
\end{aligned}
\label{eq:opt1}
\end{equation}
Otherwise, if the original-data-transmission condition is satisfied for all 
devices, the edge server reconstructs the sensing matrices and sets 
$U_k=X_k$. The optimization problem then reduces to the raw-data 
multi-view inference problem
\begin{equation}
\textbf{(P2)}\quad
\min\ \mathrm{H}\left(Y\mid X_1,\ldots,X_K\right).
\label{eq:opt2}
\end{equation} 

We next analyze problem~\textbf{P1} for task-oriented transmission through a variational approximation and decompose it into two separately trained subproblems: adaptive distributed encoding (ADE) at the devices and multi-view inference (MI) at the edge server. The MI formulation is also applicable to~\textbf{P2}, with the server-side input changed from the uploaded representations $\{\mathbf{Z}_k\}_{k=1}^{K}$ to the reconstructed raw sensing matrices $\{\mathbf{X}_k\}_{k=1}^{K}$. The remainder of this section first presents the problem analysis, followed by the ADE and MI solutions. 





\subsection{Problem Analysis}
\label{subsec:lossy_variational}

This subsection focuses on how to approximate
the objective function in \textbf{P1} using variational distribution to enable distributed encoding and multi-view inference. 
Recall that the local encoder at device $k$ is modeled by
$p_{\theta_k}(\mathbf{Z}_k\mid\mathbf{X}_k)$, while the multi-view inference model at the edge server is modeled by
$p_{\psi}(y\mid\mathbf{Z}_1,\ldots,\mathbf{Z}_K)$. The conditional entropy in \eqref{eq:opt1} can be upper bounded using a variational predictive distribution as
\begin{align}
    \mathrm{H}(Y|Z_1,\cdots,Z_K) \leq  - \mathbb{E}_{p(y, \mathbf{Z}_1,\cdots, \mathbf{Z}_K)}\left[\log p_{\psi}(y | \mathbf{Z}_1, \ldots, \mathbf{Z}_K) \right], \label{eq:con_entropy}
\end{align}
Given the Markov chain $Y \leftrightarrow X_k \leftrightarrow Z_k$,  the joint distribution $p(y, \mathbf{Z}_1,\cdots, \mathbf{Z}_K)$ can be factorized as
\begin{align}\label{eq: joint dis}
&p(y, \mathbf{Z}_1, \ldots, \mathbf{Z}_K) \nn \\
&=p(y) \int 
p(\mathbf{X}_1,\ldots,\mathbf{X}_K \mid y)
\prod_{k=1}^K 
p_{\theta_k}(\mathbf{Z}_k \mid \mathbf{X}_k)
\, d\mathbf{X}_1 \cdots d\mathbf{X}_K.
\end{align}
Substituting \eqref{eq: joint dis} into \eqref{eq:con_entropy}, we have 
\begin{align}
    &\mathrm{H}(Y|Z_1,\cdots, Z_K)\nn\\ 
   &\leq - \int p(y)
\int \int
p(\mathbf{X}_1,\ldots,\mathbf{X}_K \mid y)
\prod_{k=1}^K
p_{\theta_k}(\mathbf{Z}_k \mid \mathbf{X}_k)
\nn\\
&\qquad \qquad \qquad  \times
\log p_\psi
(y \mid \mathbf{Z}_1,\ldots,\mathbf{Z}_K)
\, d\mathbf{X}_{1:K} \,
d\mathbf{Z}_{1:K} \, dy  \nn \\
    &\approx -\frac{1}{L}\sum_{l=1}^L  \log p_{\psi}(y^{(l)}|\mathbf{Z}_1^{(l)},\cdots, \mathbf{Z}_K^{(l)}), \label{eq: obj_sampling}
\end{align} 
where the final step uses a Monte Carlo approximation with $L$ samples.
Specifically, given a paired multi-view training sample
$(\mathbf{X}_1^{(l)},\ldots,\mathbf{X}_K^{(l)},y^{(l)})$
drawn from the joint data distribution $p(\mathbf{X}_1,\ldots,\mathbf{X}_K \mid y)$, we sample
$\mathbf{Z}_k^{(l)}$ from
$p_{\theta_k}(\mathbf{Z}_k \mid \mathbf{X}_k^{(l)})$
for each device $k$. This process is repeated for $l=1,\cdots, L$, and the resulting samples are used to approximate the expectation integral in the objective function.

Minimizing \eqref{eq: obj_sampling} requires joint updates $\{\theta_k\}_{k=1}^K$ and $\psi$ across all devices and the edge server. This process involves frequent communication between the server and devices to exchange updates of the encoded features $\{{\bf Z}_1^{(l)}, \cdots, {\bf Z}_K^{(l)}\}_{l=1}^L$ and synchronize parameters during each iteration. Such an approach is communication-intensive and poses significant challenges for distributed systems. To address this, we propose decomposing the optimization problem {\bf P1} in \eqref{eq:opt1} into two subproblems: 
\begin{equation}
\textbf{(P1.1)} \quad
\begin{aligned} \label{eq:opt1-1} 
    &\min \quad \mathrm{H}(Y| Z_k) \\
     &\mathrm{s.t.}  \quad \mathrm{I}({X}_k;{Z}_k)\leq C_k\tau_{\mathrm{req}}. 
\end{aligned}
\end{equation}
which can be solved independently on each device $k$. Once solved, each device uses its well-trained encoder to generate  $\{\mathbf{Z}_k^{(l)}\}_{l=1}^L$, and uploads them to the edge server in a one-shot transmission. The edge server then solves the multiview inference problem:
\vspace{-1.mm}
\begin{equation}
\textbf{(P1.2)} \quad
\begin{aligned} \label{eq:opt1-2} 
    &\min \quad \mathrm{H}(Y|Z_1,\cdots, Z_K) 
\end{aligned}
\end{equation} 
using the received encoding vectors $\{\mathbf{Z}_1^{(l)}, \cdots, \mathbf{Z}_K^{(l)}\}_{l=1}^L$ and target labels $\{y^{(l)}\}_{l=1}^L$.

This decomposition enables a two-stage optimization process, termed adaptive distributed encoding and multi-view inference (ADE-MI). In ADE, each device independently trains a local encoder to extract task-relevant representations while satisfying its individual communication constraint. This is equivalent to minimizing an upper bound of {\bf P1}, as we have $\mathrm{H}(Y\mid Z_1,\ldots,Z_K) \leq \min_k \mathrm{H}(Y\mid Z_k)$.  
In MI, the edge server trains a joint inference model using the available multi-view inputs. Problems~\textbf{P1.2} and~\textbf{P2} share the same MI formulation, 
using~$\{\mathbf{Z}_k\}_{k=1}^{K}$ in the task-oriented case and $\{\mathbf{X}_k\}_{k=1}^{K}$ in the lossless case as the server-side inputs, respectively. The following subsections develop the ADE and MI solutions.

\subsection{Adaptive Distributed Encoding (ADE)}
\label{subsec:ade}
To address the communication constraint in terms of bit rate, the local encoder $\theta_k$ at device $k$ consists of a trainable feature extractor followed by a uniform quantization module. We adopt uniform quantization due to its simplicity in calculating the bit length. Each element in $\mathbf{Z}_k \in \mathbb{R}^{d_k}$ is  discretized into $n_k$ bits, therefore, the number of encoded bits to be transmitted is $d_k n_k$. The channel capacity $C_k$ together with latency requirement $\tau_{\mathrm{req}}$ regulate the maximum allowable bit-rate for device $k$. We fix $d_k$ while adapting $n_k$ to meet communication budget, whilst avoiding changes in the encoder output dimension that would otherwise require separate model training for different communication conditions. Specifically, under the communication constraint in problem~\textbf{(P1.1)}, the number of quantization bits is selected as
\begin{equation}n_k=\left\lfloor\frac{C_k\tau_{\mathrm{req}}}{d_k}\right\rfloor.\label{eq:quantization_bits}
\end{equation}

With the communication constraint satisfied by~\eqref{eq:quantization_bits}, we next minimize the objective function~$\mathrm{H}(Y\mid Z_k)$ in the problem~\textbf{P1.1}. Following a similar approach to that in \eqref{eq: obj_sampling}, for device $k$, we introduce an auxiliary local classifier
$p_{\omega_k}(y\mid\mathbf{Z}_k)$, parameterized by $\omega_k$,  as a 
device-side predictive distribution over the target label. 
 With $\widetilde{\mathbf{Z}}_k$ as the unquantized version of $\mathbf{Z}_k$, we model its distribution $p_{\theta_k} ( \widetilde{\mathbf{Z}}_k|\mathbf{X}_k)$ as a multivariate Gaussian with mean $\boldsymbol{\mu}(\mathbf{X}_k; {\theta_k}) \in \mathbb{R}^{d_k}$ and diagonal covariance $ \boldsymbol{\sigma}(\mathbf{X}_k; {\theta_k}) \in \mathbb{R}^{d_k}$.  Both $\boldsymbol{\mu}$ and $\boldsymbol{\sigma}$  are outputs of a neural network parameterized by ${\theta_k}$ with input $\mathbf{X}_k$. Using the reparameterization trick, we sample $\widetilde{\mathbf{Z}}_k$ and then obtain $\mathbf{Z}_k$ via quantization as follows:  \vspace{-0.5mm}
\begin{align} 
\widetilde{\mathbf{Z}}_k &= \boldsymbol{\mu}(\mathbf{X}_k; {\theta_k})  + \boldsymbol{\sigma}(\mathbf{X}_k; {\theta_k})  \odot \boldsymbol{\epsilon}  \label{eq:reparamete}   \\ 
\mathbf{Z}_k &= \mathcal{Q}(\widetilde{\mathbf{Z}}_k), \label{eq:discrete}
\end{align}
where $\boldsymbol{\epsilon} \sim \mathcal{N}(\mathbf{0},\mathbf{I})$, $\odot$ denotes element-wise product, and $\mathcal{Q}(\cdot)$ represents the uniform quantization.
Given a batch of data $\{(\mathbf{X}_k^{(l)}, y^{(l)})\}_{l=1}^L$, we approximate the objective function $\mathrm{H}(Y| Z_k)$  using Monte Carlo sampling as
\begin{equation}
\mathcal{L}_{\text{device},k}(\theta_k,\omega_k) \simeq  - \frac{1}{L}\sum_{l=1}^L \log p_{\omega_k}(y^{(l)}|\mathbf{Z}_k^{(l)}).\label{eq:loss1}
\end{equation}
The optimization problem {\bf P1.1} can then be solved by minimizing \eqref{eq:loss1} with respect to $\theta_k$ and $\omega_k$ as described in Algorithm~1. The constraint is satisfied by adapting the quantization precision of $\mathbf{Z}_k$ according to \eqref{eq:quantization_bits}.
After the local encoders have been trained, their parameters
${\theta_k}$ are fixed. The edge server then trains the
multi-view inference model using the encoded representations received
from all devices as described next.

\subsection{Multi-View Inference (MI)}
\label{subsec:lossless_inference}
For task-oriented transmission, during training, upon receiving the encoded features 
$\{{\bf Z}_1^{(l)}, \cdots, {\bf Z}_K^{(l)}\}_{l=1}^L$ 
and their associated labels $\{y^{(l)}\}_{l=1}^L$, 
the predictive distribution 
$p_{\psi}(y\mid{\bf Z}_1,\ldots,{\bf Z}_K)$, parameterized by $\psi$, 
is used to approximate the conditional entropy in \eqref{eq:opt1-2}. 
Following the same Monte Carlo approximation as in \eqref{eq: obj_sampling}, 
the objective function is approximated as:  
\begin{equation}
\mathcal{L}_{\text{server}}(\psi) \simeq \frac{1}{L} \sum_{l=1}^L -\log p_{\psi}(y^{(l)}|{\bf Z}_1^{(l)},\cdots, {\bf Z}_K^{(l)}).\label{eq:loss2}
\end{equation} 
Similarly, for original-data transmission, given the batch of received sensing data-label pairs $\{{\bf X}_1^{(l)}, \cdots, {\bf X}_K^{(l)}, y^{(l)}\}_{l=1}^L$, and the inference model $\phi$,  the objective in~\eqref{eq:opt2} is approximated~as

\begin{equation}
\mathcal{L}_{\text{server}}(\phi) \simeq \frac{1}{L} \sum_{l=1}^L -\log p_{\phi}(y^{(l)}|{\bf X}_1^{(l)},\cdots ,{\bf X}_K^{(l)}).\label{eq:loss3}
\end{equation} 
The MI problem is solved by minimizing \eqref{eq:loss2} with respect to 
$\psi$ for task-oriented transmission, or by minimizing \eqref{eq:loss3} with 
respect to $\phi$ for original-data transmission, as outlined in Algorithm~2.

\begin{algorithm}[t]
\caption{On-Device Local Training for Adaptive Distributed Encoding }
\label{alg:local_training}
\KwIn{Local training dataset $\mathcal{D}_k$, learning rate $\eta_k$,
batch size $L$, number of epochs $I_k$, channel rate $C_k$,
latency requirement $\tau_{\mathrm{req}}$ and representation
dimension $d_k$}

Compute the quantization bit number $n_k$ using
\eqref{eq:quantization_bits};

Initialize parameters $\theta_k^{(0)}$
and $\omega_k^{(0)}$ according to $d_k$;

\For{$i=1$ \KwTo $I_k$}{
Randomly select a minibatch
${(\mathbf{X}_k^{(l)},y^{( l)})}_{l=1}^{L}$
from $\mathcal{D}_k$;

\For{$l=1$ \KwTo $L$}{
    Compute
    $\boldsymbol{\mu}
    (\mathbf{X}_k^{(l)};\theta_k^{(i-1)})$
    and
    $\boldsymbol{\sigma}
    (\mathbf{X}_k^{(l)};\theta_k^{(i-1)})$\;

    Sample
    $\boldsymbol{\epsilon}^{(l)}
    \sim\mathcal{N}(\mathbf{0},\mathbf{I})$\;

    Compute
    $\mathbf{Z}_k^{(l)}$ based on \eqref{eq:reparamete} and \eqref{eq:discrete}\;
}

Update $\theta_k$ and $\omega_k$ based on \eqref{eq:loss1}\;

$\theta_k^{(i)}=\theta_k^{(i-1)}-\eta_k
\nabla_{\theta_k}
\mathcal{L}_{\mathrm{device},k}
(\theta_k^{(i-1)},\omega_k^{(i-1)})$\;

$\omega_k^{(i)}=\omega_k^{(i-1)}-\eta_k
\nabla_{\omega_k}
\mathcal{L}_{\mathrm{device},k}
(\theta_k^{(i-1)},\omega_k^{(i-1)})$\;
}

Given $\theta_k^{(I_k)}$, sample $\mathbf{Z}_k^{(l)}$ for $L$ times based on \eqref{eq:reparamete} and \eqref{eq:discrete}\;
\KwOut{$\{\mathbf{Z}_k^{(l)}\}_{l=1}^L$ and optimized parameters $\theta_k^{(I_k)}$}
\end{algorithm}

\begin{algorithm}[t]
\caption{Multi-View Inference Models Training at the Edge Server}
\label{alg:server_training}
\KwIn{ Encoded feature-label pairs $\{{\bf Z}_1^{(l)},\cdots,{\bf Z}_K^{(l)},y^{(l)}\}_{l=1}^L$, sensing data-label pairs $\{{\bf X}_1^{(l)},\cdots, {\bf X}_K^{(l)},y^{(l)}\}_{l=1}^L$, learning rate $\eta$ and number of epochs $I$}

Randomly initialize parameters: $\psi^{(0)}$ and $\phi^{(0)}$ \;

\For{$i = 1$ to $I$}{
     Update parameters $\psi$ using $\{{\bf Z}_1^{(l)},\cdots,{\bf Z}_K^{(l)},y^{(l)}\}_{l=1}^L$ based on \eqref{eq:loss2}; 
    $\psi^{(i)} = \psi^{(i-1)} - \eta\nabla_{\psi}^{(i)}\mathcal{L}_{\text{server}}(\psi^{(i-1)})$;\\
    Update parameters $\phi$ using $\{{\bf X}_1^{(l)},\cdots, {\bf X}_K^{(l)},y^{(l)}\}_{l=1}^L$ based on \eqref{eq:loss3}; 
    $\phi^{(i)} = \phi^{(i-1)} - \eta\nabla_{\phi}^{(i)}\mathcal{L}_{\text{server}}(\phi^{(i-1)})$;
}
\KwOut{Optimized parameters $\psi^{(I)}$ and $\phi^{(I)}$ }
\end{algorithm}

\section{Experimental Results}\label{Sec_exp}

We use the ``Room 1'' subset of the Widar3.0 multi-view gesture recognition dataset~\cite{9516988}, which contains 9,000 paired samples of time-series CSI measurements and gesture labels, to evaluate the proposed framework. Specifically, the samples were collected in a $2~\text{m}\times2~\text{m}$ classroom, where a Wi-Fi access point transmits signals to 6 spatially distributed receivers while a participant performs one of the 6 predefined gestures within the area. During the downlink transmission, the CSI snapshot with dimension $D$ = 121 is estimated at a frequency of 1000~Hz, resulting in a CSI packet interval of $\Delta = 1$~ms. Using the Intel-5300 WiFi chipset, each CSI element is uniformly quantized to 8 bits over $[0,1]$, yielding a CSI quantization resolution of $\delta = {1}/{255}$. 

In our experiments, unless otherwise specified, we adopt the following system parameter settings. We consider $K=3$ sensing views and a total uplink bandwidth of $B=20$~MHz, which is equally shared among the receivers, i.e., $B_k=B/K$. To meet the latency requirement of interactive Wi-Fi applications~\cite{Adame2021TSN}, we set the uplink transmission time budget to $\tau_{\mathrm{req}}=10$~ms. The observation window for each time-series CSI sample is set to $\tau = 2.018$~s, and the original CSI sequences are uniformly subsampled to evaluate the proposed framework under different CSI packet intervals, i.e., $\Delta\in\{1,4,8,16,32,64,128,256\}$~ms. For uplink channels, we use a transmission-rate efficiency factor of $\zeta=0.2$ and an SNR of $\gamma_k=10$~dB for all devices. We then generate 50 independent fading realizations for both Rayleigh and Rician channels, each consisting of 500 temporally correlated channel states, with the Rician $\mathcal{K}$-factor set to 6~dB. For ADE-MI transmission, each device outputs a 128-dimensional feature vector, i.e., $d_k=128$.



The proposed source entropy-guided adaptive transmission framework is implemented in PyTorch. For source-entropy evaluation, the NF model is constructed using autoregressive flows with causal masks \cite{papamakarios2017masked}. For adaptive transmission, the local encoders at the receivers are implemented using Transformers \cite{vaswani2017attention} to accommodate variable-length CSI sequences, while their corresponding local decoders are realized by multilayer perceptrons (MLPs). The global decoder at the Wi-Fi access point is also implemented as an MLP, taking all uploaded local features as input. For original data transmission, the decoder combines Transformer layers with an MLP-based classifier head. The dataset is split into training and testing sets at a ratio of 9:1. All models are trained for 100 epochs with a batch size of 64. We employ the adaptive moment estimation with decoupled weight decay (AdamW) optimizer, with a learning rate of $3\times10^{-4}$ and a weight decay coefficient of $10^{-4}$.

To evaluate the proposed ADE-MI method, we compare it with two representative benchmarks. The first benchmark corresponds to a task-agnostic compression scheme, where the source observations are compressed without exploiting task-specific information and transmitted under the same communication budget as ADE-MI.

\noindent $\bullet$ \textbf{Principal component analysis (PCA):} Each device first flattens the CSI sequence into a vector and projects it onto a PCA basis estimated from the training set. We retain the top $512$ components to reduce computational cost. When the communication budget $C_k\tau_{\mathrm{req}}<512$ bits, the device transmits the first $\lfloor C_k\tau_{\mathrm{req}}\rfloor$ components using $1$-bit quantization. Otherwise, it transmits all $512$ components, each quantized with $\left\lfloor C_k\tau_{\mathrm{req}}/512\right\rfloor$ bits to fully utilize the available budget. The components received at the Wi-Fi access point are fed into an MLP-based global decoder for classification.

We further consider a task-oriented compression benchmark that learns task-relevant representations for edge inference and adapts to the available communication budget.


\noindent $\bullet$ \textbf{Variational distributed deterministic information bottleneck (VDDIB)}~\cite{9837474}: The method in \cite{9837474} encodes the original inputs into multiple features and transmits them to the edge server one at a time, stopping once the inference result satisfies the required criterion. To adapt VDDIB to our setting, we modify it to transmit as many features as possible in a single transmission within the available communication budget. Specifically, we use a Transformer backbone followed by 16 independent MLP feature heads to encode the original CSI sequence into $16$ different feature representations, each of dimension $128$. When the communication budget $C_k\tau_{\mathrm{req}} < 2048$ bits, the device transmits the first $\left\lfloor C_k\tau_{\mathrm{req}}/128 \right\rfloor$ features using 1-bit quantization. Otherwise, it transmits all $16$ features, with each feature element quantized using $\left\lfloor C_k\tau_{\mathrm{req}}/2048 \right\rfloor$ bits to fully utilize the available budget. The received features are fed into an MLP-based global decoder at the edge server for classification. This implementation enables a fair comparison with the proposed ADE-MI scheme under the same communication budget constraint.

\subsection{Validation of the Source Entropy Bound}
\begin{figure}[t]
    \centering
    \includegraphics[width=0.45\textwidth]{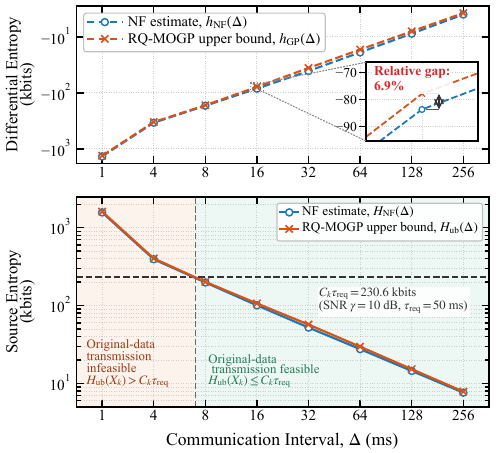}
    \caption{Validation of the proposed source-entropy analysis and the resulting original-data transmission feasibility versus communication interval $\Delta$ under $\tau_{\mathrm{req}}=50$~ms. }
    \label{fig:source_entropy}
\end{figure}

Fig.~\ref{fig:source_entropy} validates the proposed source-entropy analysis by comparing the analytical result of RQ-MOGP with the numerical estimate based on NF. As shown in the upper subplot, the analytical differential entropy closely follows the NF estimate across the evaluated communication intervals, with the inset showing a representative relative gap of $6.9\%$ at communication interval $\Delta=16$ ms. The 
RQ-MOGP result is slightly higher than the NF estimate, which verifies its role as a Gaussian-surrogate upper bound. After incorporating the quantization term, the resulting source-entropy estimates in the lower subplot remain closely aligned and decrease substantially as $\Delta$ increases, since fewer CSI observations are acquired within the fixed observation window. Comparing the available transmission budget with the RQ-MOGP source-entropy upper bound shows that original-data transmission is infeasible at $\Delta\in\{1,4\}$~ms but becomes feasible for $\Delta\geq8$~ms among the evaluated intervals. This transition supports the proposed entropy-guided criterion for switching between task-oriented and original-data transmission.

\subsection{Effectiveness of ADE-MI} 
\begin{figure}[t]
    \centering
    \includegraphics[width=0.44\textwidth]{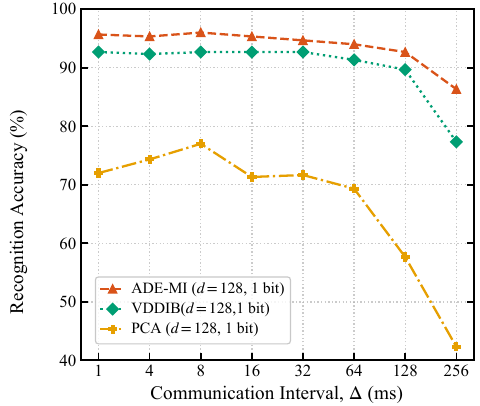}
    \caption{The impact of communication interval $\Delta$ on recognition accuracy under the same communication budget of 128 bits. }
    \label{fig:raw_vs_lossy}
\end{figure}
Fig.~\ref{fig:raw_vs_lossy} evaluates the effectiveness of the proposed ADE-MI by comparing its performance with those of VDDIB and PCA under $1$-bit feature quantization and an identical communication budget of $128$ bits. It is observed that ADE-MI consistently outperforms both baselines across all evaluated communication intervals, achieving accuracy gains of up to approximately $9.30\%$ and $44.20\%$ over VDDIB and PCA, respectively. This improvement demonstrates that ADE-MI can more effectively preserve task-relevant information under aggressive feature quantization. Moreover, as the communication interval increases from $64$ to $256$~ms, all methods exhibit declining recognition accuracy due to fewer sensing observations within the fixed observation window. Nevertheless, ADE-MI experiences a much smaller performance degradation than the baselines, demonstrating its greater robustness to sparse sensing conditions.

\begin{table}[t]
\centering
\caption{Recognition accuracy of ADE-MI with separate
and shared encoders across communication intervals}
\label{tab:multi_interval_ablation}
\scriptsize
\setlength{\tabcolsep}{2pt}
\begin{tabular}{@{}lccc@{}}
\toprule
Encoder configuration &
\makecell{No. of $\Delta$ values\\used to train\\each encoder} &
\makecell{Encoders\\per receiver} &
\makecell{Average recognition\\accuracy (\%) $\uparrow$} \\
\midrule
Separate encoders & 1 & 8 & 94.35 \\
\textbf{Shared encoder}    & 8 & 1 & \textbf{94.96} \\
\bottomrule
\end{tabular}
\end{table}
Table~\ref{tab:multi_interval_ablation} examines whether a single ADE-MI encoder can be shared across different communication intervals. Specifically, we train eight separate Transformer-based encoders, each using data from one communication interval, and then compare them with a shared encoder jointly trained on data from all eight communication intervals. At each communication interval $\Delta$, both configurations are evaluated on identical test sequences. It is observed that the shared encoder achieves an average recognition accuracy of $94.96\%$, which is comparable to, and slightly higher than, the $94.35\%$ achieved by the separate encoders. This result indicates that jointly optimizing a single set of parameters across both dense and sparse observations does not compromise recognition performance at individual communication intervals. It therefore justifies our design choice of using a single ADE-MI encoder per receiver instead of maintaining eight interval-specific encoders.

\subsection{Ablation Studies}

\begin{table}[t]
\centering
\caption{Ablation study of source-entropy-guided adaptive
transmission under time-varying Rician fading}
\label{tab:online_adaptation_ablation}
\footnotesize
\setlength{\tabcolsep}{3pt}
\renewcommand{\arraystretch}{1.1}
\begin{tabular}{@{}ccccc@{}}
\toprule
\makecell{Adaptive feature\\ quantization}
& \makecell{Source entropy-\\guided switching}
& \makecell{Average\\recognition\\accuracy (\%) $\uparrow$}
& \makecell{Outage\\(\%) $\downarrow$}
& \makecell{Original-data\\transmission\\(\%)} \\
\midrule

$\times$ 
($n_k=16$)& $\times$ & 58.91 & 37.50 & -- \\

$\times$ 
($n_k=16$)& $\checkmark$ & 59.94 & 37.50 & 34.21 \\

$\checkmark$
& $\times$ & 91.14 & 3.28 & -- \\

$\checkmark$
& $\checkmark$ & 92.23 & 3.28 & 34.21 \\
\bottomrule
\end{tabular}
\end{table}



Table~\ref{tab:online_adaptation_ablation} presents an ablation study of adaptive feature quantization and source entropy-guided switching mechanism under time-varying Rician fading channels. Without either mechanism, fixing $n_k=16$ achieves an average recognition accuracy of $58.91\%$ with an outage probability of $37.50\%$, where an outage occurs if at least one receiver cannot transmit the required bits within  $\tau_{\mathrm{req}}$. The outage state is assigned zero recognition accuracy in the evaluation.  This high outage probability indicates that fixed 16-bit quantization frequently exceeds the available transmission budget. Further introducing source-entropy-guided switching under the same fixed $n_k=16$ setting increases the recognition accuracy slightly from 58.91\% to 59.94\%, while leaving the outage probability unchanged at 37.50\%. The limited improvement shows that source-entropy-guided switching alone cannot compensate for the accuracy loss caused by a fixed transmission configuration.

By contrast, enabling adaptive $n_k$ without switching reduces the outage probability to 3.28\% and improves the recognition accuracy to 91.14\%, demonstrating that adapting the transmission configuration to the channel condition is essential for maintaining reliable inference under time-varying Rician channels. When source-entropy-guided switching is further incorporated, the recognition accuracy increases to 92.23\% without any additional outage penalty, while original-data transmission is selected in 34.21\% of the channel states. This reveals that the two mechanisms are complementary: adaptive transmission primarily mitigates transmission outages by dynamically adjusting $n_k$ to the time-varying channel conditions, whereas source-entropy-guided switching further improves recognition performance by employing original-data transmission when available.

\subsection{Impact of Channel SNR}

\begin{figure}[t]
    \centering
    \includegraphics[width=0.48\textwidth]{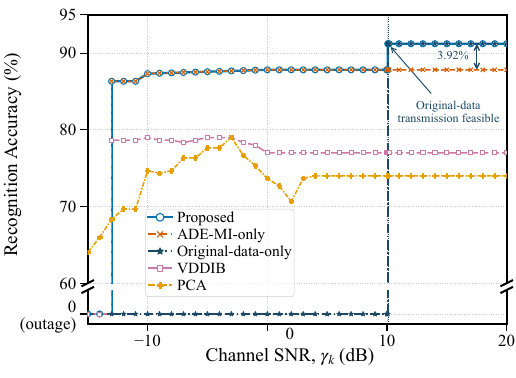}
    \caption{The impact of channel SNR $\gamma_k$ on recognition accuracy at $\Delta=256$~ms and $\tau_{\mathrm{req}}=2$~ms.}
    \label{fig:static_rate_performance}
\end{figure}

Fig.~\ref{fig:static_rate_performance} evaluates the impact of channel SNR $\gamma_k$ on recognition accuracy at $\Delta=256$~ms and $\tau_{\mathrm{req}}=2$~ms. It is observed that the proposed scheme maintains high recognition accuracy across a wide range of channel conditions whenever transmission is feasible. In particular, when $\gamma_k$ reaches approximately $10$~dB, the transmission budget $C_k\tau_{\mathrm{req}}$ becomes sufficient for original-data transmission at all receivers. The proposed scheme accordingly switches to original-data transmission, achieving an accuracy gain of approximately $3.92\%$ over ADE-MI-only and attaining the same performance as original-data-only. Below this threshold, original-data-only remains in outage, whereas the proposed scheme employs ADE-MI and outperforms VDDIB and PCA whenever ADE-MI transmission is feasible. At very low SNRs, the proposed scheme, ADE-MI-only, and VDDIB also enter outage, while PCA remains feasible by transmitting fewer principal components. These results demonstrate that the proposed adaptive scheme can effectively maintain recognition performance under limited communication capacity and further improve accuracy by exploiting original-data transmission when sufficient communication resources are available.

\begin{figure}[t]
    \centering
    \includegraphics[width=0.44\textwidth]{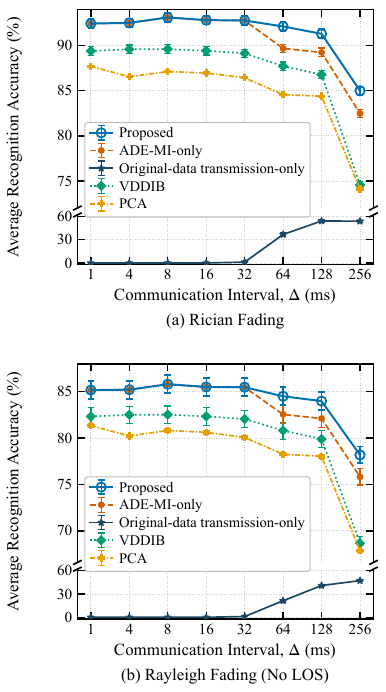}
    \caption{The impact of communication interval $\Delta$ on average recognition
    accuracy under time-varying fading channels. }
    \label{fig:dynamic_switching}
\end{figure}

\subsection{Impact of Communication Interval} 
Fig.~\ref{fig:dynamic_switching} evaluates the impact of communication interval~$\Delta$ on average recognition accuracy under time-varying Rician and Rayleigh fading channels. It is observed in Fig.~\ref{fig:dynamic_switching}(a) and (b) that the proposed scheme achieves the highest average accuracy under both fading conditions and outperforms ADE-MI-only when $\Delta \geq 64$~ms. This is because a larger $\Delta$ results in fewer CSI observations within the fixed sensing window, which reduces the estimated number of bits required for original-data transmission. The original-data transmission becomes feasible in a larger fraction of channel states, and the proposed scheme can therefore use it more frequently to maintain recognition accuracy. Nevertheless, the accuracy of all schemes decreases significantly at $\Delta=256$~ms, which suggests that too few sensing observations may limit recognition performance when the communication interval becomes too large. Moreover, all schemes achieve lower average accuracy under Rayleigh fading than under Rician fading, because the absence of a line-of-sight (LOS) component results in more severe channel fluctuations and less reliable transmission.

\begin{figure}[t]
    \centering
\includegraphics[width=0.45\textwidth]{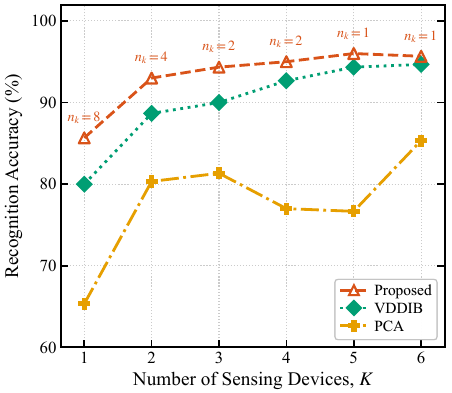}
    \caption{The impact of the number of sensing devices $K$ on average
    recognition accuracy under a fixed total uplink bandwidth $B = 20$~MHz under $\Delta = 128$~ms, $\gamma = -10$~dB, $\tau_{\text{req}}=2$~ms.}
    \label{fig:view_count_tradeoff}
\end{figure}

\subsection{Impact of the Number of Sensing Devices}

Fig.~\ref{fig:view_count_tradeoff} evaluates the trade-off between multi-view information gain and the reduced per-device communication budget as the number of sensing devices $K$ increases under a fixed total uplink bandwidth. It is observed that the proposed scheme achieves the highest accuracy for all $K$ and benefits from additional sensing views despite the decreasing per-device communication budget. Specifically, as $K$ increases from 1 to 5, our scheme decreases the feature quantization precision $n_k$ from 8 to 1 to adapt to the varying communication budget, whereas the corresponding recognition accuracy increases from approximately $86\%$ to $96\%$. This indicates that the gain from additional sensing views outweighs the loss caused by coarser quantization at each device. By contrast, PCA achieves lower accuracy and even decreases from approximately $81\%$ at $K=3$ to $77\%$ at $K=5$, although its accuracy recovers at $K=6$. These results demonstrate that the proposed task-oriented encoding can effectively balance multi-view information gain and per-device communication constraints.

 


\section{Conclusion}\label{Sec5}

In this paper, we developed a source entropy-guided adaptive transmission framework for communication-driven multi-view sensing. We characterized the entropy of packet-triggered CSI as a function of the communication interval and derived a tractable analytical bound that enables the sensing information demand to be matched with the available uplink bit budget. Based on this relationship, the framework adapts between original-data and task-oriented transmission, and we developed ADE-MI to enable communication-efficient multi-view inference in the latter case.  Experiments validated the entropy analysis and demonstrated the effectiveness of the adaptive framework under time-varying channel conditions.

Future work can extend the framework to joint view selection and bandwidth allocation, where the number of participating devices and their bandwidth shares are optimized according to their source information and channel conditions. 
Another direction is to allow devices to select transmission modes independently and investigate how the edge server should fuse heterogeneous multi-view inputs consisting of both original data and task-oriented features.  A possible solution is to design a modality-aware encoder, similar to ADE, that maps raw data into the space aligned with the task-oriented  features before multi-view inference.

\bibliographystyle{IEEEtran}
\bibliography{mproj}

\end{document}